\documentclass[10pt, conference, letterpaper]{IEEEtran}
\usepackage{url}
\usepackage[utf8]{inputenc}
\usepackage{xcolor}
\usepackage{amsmath}
\usepackage{multirow}
\usepackage{amssymb}
\usepackage{enumitem}
\usepackage{bm}
\usepackage{graphicx}

\usepackage{tikz}
\usetikzlibrary{plotmarks,patterns,decorations.pathreplacing,backgrounds,calc,arrows,arrows.meta,spy,matrix}

\usepackage{gensymb}

\usepackage[acronyms,nonumberlist,nopostdot,nomain,nogroupskip]{glossaries}
\usepackage{tablefootnote}
\usepackage{booktabs}
\usepackage{tabularx}
\usepackage{epsfig}
\usepackage{epstopdf}

\usepackage{pgfplots}
\pgfplotsset{compat=newest} 
\pgfplotsset{plot coordinates/math parser=false} 
\newlength\fheight
\newlength\fwidth
\usepgfplotslibrary{patchplots,groupplots}
\usepackage{tikzscale}
\usepackage{siunitx}

\usepackage{multirow}

\usepackage[font=scriptsize]{subcaption}
\usepackage[font=footnotesize]{caption}

\usepackage{mathtools}

\usepackage{dblfloatfix}    
\usepackage{colortbl}

\newacronym{3gpp}{3GPP}{3rd Generation Partnership Project}
\newacronym{adc}{ADC}{Analog to Digital Converter}
\newacronym{5g}{5G}{5th generation}
\newacronym{6g}{6G}{6th generation}
\newacronym{aimd}{AIMD}{Additive Increase Multiplicative Decrease}
\newacronym{am}{AM}{Acknowledged Mode}
\newacronym{amc}{AMC}{Adaptive Modulation and Coding}
\newacronym{aqm}{AQM}{Active Queue Management}
\newacronym{awgn}{AGWN}{Additive White Gaussian Noise}
\newacronym{balia}{BALIA}{Balanced Link Adaptation}
\newacronym{bdp}{BDP}{Bandwidth-Delay Product}
\newacronym{qos}{QoS}{Quality of Service}
\newacronym{qoe}{QoE}{Quality of Experience}
\newacronym{pqos}{PQoS}{Predictive Quality of Service}
\newacronym{bf}{BF}{Beamforming}
\newacronym{cc}{CC}{Congestion Control}
\newacronym{cu}{CU}{Centralized Unit}
\newacronym{ai}{AI}{artificial intelligence}
\newacronym{dql}{DQL}{Double Q-learning}
\newacronym{du}{DU}{Distributed Unit}
\newacronym{cdf}{CDF}{Cumulative Distribution Function}
\newacronym{lidar}{LiDAR}{Light Detection and Ranging}
\newacronym{cn}{CN}{Core Network}
\newacronym{rl}{RL}{Reinforcement Learning}
\newacronym{cam}{CAM}{Cooperative Awareness Message}
\newacronym{mse}{MSE}{Mean Squared Error}
\newacronym{cqi}{CQI}{Channel Quality Information}
\newacronym{mdp}{MDP}{Markov Decision Process}
\newacronym{cp}{CP}{Control Plane}
\newacronym{csirs}{CSI-RS}{Channel State Information - Reference Signal}
\newacronym{dc}{DC}{Dual Connectivity}
\newacronym{dce}{DCE}{Direct Code Execution}
\newacronym{dci}{DCI}{Downlink Control Information}
\newacronym{dl}{DL}{Downlink}
\newacronym{dmr}{DMR}{Deadline Miss Ratio}
\newacronym{dmrs}{DMRS}{DeModulation Reference Signal}
\newacronym{e2e}{E2E}{end-to-end}
\newacronym{ecn}{ECN}{Explicit Congestion Notification}
\newacronym{edf}{EDF}{Earliest Deadline First}
\newacronym{enb}{eNB}{evolved Node Base}
\newacronym{epc}{EPC}{Evolved Packet Core}
\newacronym{es}{ES}{Edge Server}
\newacronym{fdma}{FDMA}{Frequency Division Multiple Access}
\newacronym{fdd}{FDD}{Frequency Division Duplexing}
\newacronym[firstplural=Radio Access Technologies (RATs)]{rat}{RAT}{Radio Access Technology}
\newacronym{fs}{FS}{Fast Switching}
\newacronym{ftp}{FTP}{File Transfer Protocol}
\newacronym{gnb}{gNB}{Next Generation Node Base}
\newacronym{harq}{HARQ}{Hybrid Automatic Repeat reQuest}
\newacronym{hetnet}{HetNet}{Heterogeneous Network}
\newacronym{hh}{HH}{Hard Handover}
\newacronym{hol}{HOL}{Head-of-Line}
\newacronym{ia}{IA}{Initial Access}
\newacronym{ieee}{IEEE}{Institute of Electrical and Electronics Engineers}
\newacronym{imt}{IMT}{International Mobile Telecommunication}
\newacronym{iot}{IoT}{Internet of Things}
\newacronym{ldpc}{LDPC}{Low-Density Parity Check}
\newacronym{los}{LOS}{Line-of-Sight}
\newacronym{lte}{LTE}{Long Term Evolution}
\newacronym{m2m}{M2M}{Machine to Machine}
\newacronym{ml}{ML}{machine learning}
\newacronym{mac}{MAC}{Medium Access Control}
\newacronym{mc}{MC}{Multi-Connectivity}
\newacronym{mcs}{MCS}{Modulation and Coding Scheme}
\newacronym{mec}{MEC}{Mobile Edge Cloud}
\newacronym{mi}{MI}{Mutual Information}
\newacronym{mimo}{MIMO}{Multiple Input, Multiple Output}
\newacronym{mmwave}{mmWave}{millimeter wave}
\newacronym{mptcp}{MPTCP}{Multipath TCP}
\newacronym{mr}{MR}{Maximum Rate}
\newacronym{mss}{MSS}{Maximum Segment Size}
\newacronym{mtd}{MTD}{Machine-Type Device}
\newacronym{mtu}{MTU}{Maximum Transmission Unit}
\newacronym{nfv}{NFV}{Network Function Virtualization}
\newacronym{nlos}{NLOS}{Non-Line-of-Sight}
\newacronym{nlosv}{NLOSv}{Vehicle Non-Line-of-Sight}
\newacronym{nr}{NR}{New Radio}
\newacronym{ofdm}{OFDM}{Orthogonal Frequency Division Multiplexing}
\newacronym{pdcch}{PDCCH}{Physical Downlonk Control Channel}
\newacronym{pdcp}{PDCP}{Packet Data Convergence Protocol}
\newacronym{pdsch}{PDSCH}{Physical Downlink Shared Channel}
\newacronym{pdu}{PDU}{Packet Data Unit}
\newacronym{pf}{PF}{Proportional Fair}
\newacronym{pgw}{PGW}{Packet Gateway}
\newacronym{phy}{PHY}{Physical}
\newacronym{pbch}{PBCH}{Physical Broadcast Channel}
\newacronym[plural=\gls{mme}s,firstplural=Mobility Management Entities (MMEs)]{mme}{MME}{Mobility Management Entity}
\newacronym{prb}{PRB}{Physical Resource Block}
\newacronym{pss}{PSS}{Primary Synchronization Signal}
\newacronym{pscch}{PSCCH}{Physical Sidelink Control Channel}
\newacronym{pucch}{PUCCH}{Physical Uplink Control Channel}
\newacronym{pusch}{PUSCH}{Physical Uplink Shared Channel}
\newacronym{rach}{RACH}{Random Access Channel}
\newacronym{ran}{RAN}{Radio Access Network}
\newacronym{red}{RED}{Random Early Detection}
\newacronym{rf}{RF}{Radio Frequency}
\newacronym{rlc}{RLC}{Radio Link Control}
\newacronym{rlf}{RLF}{Radio Link Failure}
\newacronym{rrc}{RRC}{Radio Resource Control}
\newacronym{rrm}{RRM}{Radio Resource Management}
\newacronym{rru}{RRU}{Remote Radio Unit}
\newacronym{rr}{RR}{Round Robin}
\newacronym{rs}{RS}{Remote Server}
\newacronym{rsrp}{RSRP}{Reference Signal Received Power}
\newacronym{rss}{RSS}{Received Signal Strength}
\newacronym{rtt}{RTT}{Round Trip Time}
\newacronym{rw}{RW}{Receive Window}
\newacronym{rx}{RX}{Receiver}
\newacronym{sa}{SA}{standalone}
\newacronym{sack}{SACK}{Selective Acknowledgment}
\newacronym{sap}{SAP}{Service Access Point}
\newacronym{sc}{SC}{Single Carrier}
\newacronym{sch}{SCH}{Secondary Cell Handover}
\newacronym{scoot}{SCOOT}{Split Cycle Offset Optimization Technique}
\newacronym{sdma}{SDMA}{Spatial Division Multiple Access}
\newacronym{sinr}{SINR}{Signal to Interference plus Noise Ratio}
\newacronym{sl}{SL}{Sidelink}
\newacronym{sm}{SM}{Saturation Mode}
\newacronym{snr}{SNR}{Signal-to-Noise-Ratio}
\newacronym{son}{SON}{Self-Organizing Network}
\newacronym{ss}{SS}{Synchronization Signal}
\newacronym{srs}{SRS}{Sounding Reference Signal}
\newacronym{sss}{SSS}{Secondary Synchronization Signal}
\newacronym{tb}{TB}{Transport Block}
\newacronym{tcp}{TCP}{Transmission Control Protocol}
\newacronym{tdd}{TDD}{Time Division Duplexing}
\newacronym{tdma}{TDMA}{Time Division Multiple Access}
\newacronym{tfl}{TfL}{Transport for London}
\newacronym{tm}{TM}{Transparent Mode}
\newacronym{trp}{TRP}{Transmitter Receiver Pair}
\newacronym{tti}{TTI}{Transmission Time Interval}
\newacronym{ttt}{TTT}{Time-to-Trigger}
\newacronym{tx}{TX}{Transmitter}
\newacronym{ue}{UE}{User Equipment}
\newacronym{ul}{UL}{Uplink}
\newacronym{uml}{UML}{Unified Modeling Language}
\newacronym{um}{UM}{Unacknowledged Mode}
\newacronym{utc}{UTC}{Urban Traffic Control}
\newacronym{vm}{VM}{Virtual Machine}
\newacronym{rsrq}{RSRQ}{Reference Signal Received Quality}
\newacronym{rssi}{RSSI}{Received Signal Strength Indicator}
\newacronym{crs}{CRS}{Cell Reference Signal}
\newacronym{nsa}{NSA}{Non Stand Alone}
\newacronym{mrdc}{MR-DC}{Multi \gls{rat} \gls{dc}}
\newacronym{endc}{EN-DC}{E-UTRAN-\gls{nr} \gls{dc}}
\newacronym{5gc}{5GC}{5G Core}
\newacronym{si}{SI}{Study Item}
\newacronym{iab}{IAB}{Integrated Access and Backhaul}
\newacronym{wf}{WF}{Wired-first}
\newacronym{hqf}{HQF}{Highest-quality-first}
\newacronym{pa}{PA}{Position-aware}
\newacronym{mlr}{MLR}{Maximum-local-rate}
\newacronym{wbf}{WBF}{Wired Bias Function}
\newacronym{mib}{MIB}{Master Information Block}
\newacronym{sib}{SIB}{Secondary Information Block}
\newacronym{rnti}{RNTI}{Radio Network Temporary Identifier}
\newacronym{dft}{DFT}{Discrete Fourier Transform}
\newacronym{kpi}{KPI}{Key Performance Indicator}
\newacronym{ppp}{PPP}{Poisson Point Process}
\newacronym{v2v}{V2V}{Vehicle-to-Vehicle}
\newacronym{wave}{WAVE}{Wireless Access in Vehicular Environments}
\newacronym{udp}{UDP}{User Datagram Protocol}
\newacronym{upa}{UPA}{Uniform Planar Array}
\newacronym{fec}{FEC}{Forward Error Correction}
\newacronym{v2x}{V2X}{Vehicle-To-Everything}
\newacronym{psfch}{PSFCH}{Physical Sidelink Feedback Channel}
\newacronym{pssch}{PSSCH}{Physical Sidelink Shared Channel}
\newacronym{csma}{CSMA}{Carrier Sense Multiple Access}
\newacronym{v2n}{V2N}{Vehicle-to-Network}
\newacronym{wlan}{WLAN}{Wireless Local Area Network}
\newacronym{cav}{CAV}{Connected and Autonomous Vehicle}
\newacronym{v2i}{V2I}{Vehicle-to-Infrastructure}
\newacronym{d2d}{D2D}{Device-to-Device}
\newacronym{c-its}{C-ITS}{Connected Intelligent Transportation System}
\newacronym{fr2}{FR2}{Frequency Range 2}
\newacronym{fr1}{FR1}{Frequency Range 1}
\newacronym{bs}{BS}{Base Station}
\newacronym{sdu}{SDU}{Service Data Unit}
\newacronym{csi}{CSI}{Channel State Information}
\newacronym{scs}{SCS}{Subcarrier Spacing}
\newacronym{sumo}{SUMO}{Simulation of Urban MObility}
\newacronym{prr}{PRR}{Packet Reception Ratio}
\newacronym{dnn}{DNN}{Deep Neural Network}
\newacronym{pdr}{PDR}{Packet Delivery Ratio}
\newacronym{edca}{EDCA}{Enhanced Distribution Channel Access}
\newacronym{sdap}{SDAP}{Service Data Adaptation Protocol}
\newacronym{osm}{OSM}{OpenStreetMap}
\newacronym{rsu}{RSU}{Road Side Unit}
\newacronym{hv}{HV}{Host Vehicle}
\newacronym{imsi}{IMSI}{International Mobile Subscriber Identity}
\newacronym{lcid}{LCID}{Logical Channel Identifier}
\newacronym{nnet}{NN}{Neural Network}
\newacronym{movav}{MovAv}{Moving Average}
\newacronym{idw}{IDW}{Inverse Distance Weighting}
\newacronym{lin}{Lin}{Linear Interpolation}
\newacronym{relu}{ReLU}{Rectified Linear Unit}
\newacronym{adam}{Adam}{Adaptive moment estimator}
\newacronym{drl}{DRL}{Deep Reinforcement Learning}
\newacronym{fcd}{FCD}{Floating Car Data}
\newacronym{csv}{CSV}{Comma Separated Values}
\newacronym{ca}{CA}{Carrier Aggregation}
\newacronym{nwdaf}{NWDAF}{Network Data Analytics Function}
\newacronym{etsi}{ETSI}{European Telecommunications Standards Institute}

\definecolor{steelblue}{RGB}{176,196,222}

\usepackage{hyperref}
\usepackage[capitalize]{cleveref}

\begin{document}


\title{A Reinforcement Learning Framework for PQoS \\ in a Teleoperated Driving Scenario}

\author{\IEEEauthorblockN{{Federico Mason$^{\star}$, Matteo Drago$^{\star}$, Tommaso Zugno$^{\dagger\star}$, Marco Giordani$^{\star}$, Mate Boban$^{\dagger}$, Michele Zorzi$^{\star}$}\medskip}
\IEEEauthorblockA{
$^{\star}$Department of Information Engineering, University of Padova, Italy. Email: \texttt{\{name.surname\}@dei.unipd.it}\\
$^{\dagger}$Huawei Technologies, Munich Research Center, Germany. Email: \texttt{name.surname@huawei.com}}\\[-3.5ex]}

\maketitle

\IEEEaftertitletext{\vspace{-3ex}}

\begin{abstract}
In recent years, autonomous networks have been designed with \gls{pqos} in mind, as a means for applications operating in the industrial and/or automotive sectors to predict unanticipated \gls{qos} changes and react accordingly. In this context, \gls{rl} has come out as a promising approach to perform accurate predictions, and optimize the efficiency and adaptability of wireless networks. Along these lines, in this paper we propose the design of a new entity, integrated at the \acrshort{ran} level that implements PQoS functionalities with the support of an RL framework. Specifically, we focus on the design of the reward function of the learning agent, able to convert QoS estimates into appropriate countermeasures if QoS requirements are not satisfied. We demonstrate via~\text{ns-3} simulations that our approach achieves better results in terms of \gls{qos} and \gls{qoe} performance of end users in a teleoperated driving scenario.
\end{abstract}

\begin{tikzpicture}[remember picture,overlay]
\node[anchor=north,yshift=-20pt] at (current page.north) {\parbox{\dimexpr\textwidth-\fboxsep-\fboxrule\relax}{
\centering\footnotesize This paper has been submitted to IEEE WCNC 2022. Copyright may change without notice.}};
\end{tikzpicture}

\glsresetall

\section{Introduction}
\label{sec:intro}

We are witnessing a rapid evolution towards autonomous systems, able to safely operate without human intervention~\cite{shariatmadari2015machine}. In this scenario, the dynamic nature of the environment in which autonomous machines are deployed may result in the \gls{qos} to change and degrade unexpectedly, with potentially catastrophic consequences if communication failures are not promptly notified.
To solve this issue, the research community has been investigating \gls{pqos}~\cite{boban2021predictive} as a means to provide autonomous systems with advance notifications about  \gls{qos} changes. 
Unlike \emph{reactive} mechanisms, which respond to QoS deterioration only after it occurs, a \emph{predictive} approach gives time to the applications to foresee unanticipated events, and adapt their operations accordingly.
In the automotive sector, \gls{v2x} is associated to strict \gls{qos} requirements in terms of ultralow latency, as well as ultrahigh throughput, reliability, and security~\cite{giordani2020toward}. Hence, PQoS may assist \gls{v2x} networks to preemptively evaluate the potential risks of QoS degradation, and guarantee reliable~driving.

At first, network prediction was developed based on linear regression and/or filter-based models (e.g., Kalman filters)~\cite{Moral:1996,Wan:2000}.
However, linear regression came out as a good estimator when the underlying relationship between the system's variables and the response was known to be linear, even though most real systems are non-linear.
At the same time, filter-based prediction mechanisms have been shown to be very sensitive to imperfections of the environment, as well as random dynamics, and mainly focus on real-time or short-time estimation of the target state. In turn, autonomous applications may also require predictions of the system's future behavior~\cite{mason2020adaptive}, which may span several minutes or even hours~\cite{5gaa2019qos}.
Moreover, despite their quick convergence in dynamic processes, these methods tend to require a priori knowledge of the process to learn, which is not always available in autonomous systems~\cite{carron2016machine}.
More recently, \gls{ml} technologies have emerged as a powerful approach to predict and optimize wireless networks~\cite{wang2017machine}, without requiring explicitly pre-programmed a priori rules, which makes these techniques particularly interesting for enabling PQoS.

Based on the above introduction, this paper addresses the problem of developing and testing \gls{pqos} mechanisms for \gls{v2x} networks. 
To this aim, we propose the design of a new entity called ``RAN-AI'' (Sec.~\ref{sec:pqos-framework}), connected to the different components of the \gls{ran} that: (i) collects data from different sources; (ii) makes QoS predictions based on the acquired data, recognizes and notifies upcoming QoS variations; and (iii) defines and applies network countermeasures in case QoS requirements are not satisfied.
The RAN-AI entity integrates a \gls{rl} framework, able to identify the optimal (set of) countermeasures for PQoS (Sec.~\ref{sub:rl_for_pqos}). In particular, we focus on the design of the learning algorithm and the reward function of the learning agent. 
The performance of the \gls{rl} model for PQoS is validated via ns-3 simulations in a teleoperated driving scenario, considering realistic setup and input parameters (Sec.~\ref{sec:performance_evaluation}). 
We demonstrate that our RL framework guarantees the optimal trade-off between \gls{qos} and \gls{qoe} for end users (and satisfies QoS requirements of V2X applications) compared to other baseline solutions.
Finally,  we formalize our conclusions and next research steps (Sec.~\ref{sec:conclusions_and_future_works}).

\section{Our \acrshort{pqos} Framework}
\label{sec:pqos-framework}

We consider a teleoperated driving scenario, in which connected vehicles are controlled by a remote driver (either human or software). 
In view of the highly dynamic nature of the \gls{v2x} environment, PQoS shall be able to predict QoS degradation and gracefully change the operational mode of the system to ensure that end users satisfy their strict latency/reliability constraints. For example, PQoS may assist the control center in adapting the vehicles' speed and trajectory based on the condition of the radio environment~\cite{boban2021predictive}.
PQoS procedures can be implemented at both the \gls{cn} and the \gls{ran}. 
In this work we focus on the latter, and describe the functionalities of a new entity called ``RAN-AI'' that we designed to facilitate PQoS in \gls{v2x}~systems.

\subsection{RAN-AI Entity for PQoS}
\gls{ran} operations in \gls{v2x}, from modulation and coding to resource allocation and scheduling, can affect \gls{qos} performance, if not properly configured.
Along these lines, we propose the design of a RAN-AI entity, installed in a remote/edge server, or at the gNB, that executes methods to improve the network performance if the agreed QoS is not satisfied. 
In our implementation, the RAN-AI entity is connected to the different components of the RAN (\gls{cu}, \glspl{du} and \glspl{rru}), as well as to the core network by means of dedicated interfaces, and hosts an \gls{rl} agent for PQoS.  
Notably, the RAN-AI entity is in charge of the following tasks:
\begin{itemize}
  \item[(i)] Get application statistics, and collect full-stack RAN measurements from the gNB and the end users, that may be used as input parameters of the \gls{rl} agent (Sec.~\ref{sub:pqos_inputs}).
  \item[(ii)] Share such measurements with the \gls{rl} agent, which determines the optimal set of countermeasures to maximize the overall performance (Sec.~\ref{sub:pqos_countermeasures}).
  \item[(iii)] Inform the end users about QoS changes, and suggest the countermeasure(s) to adopt, based on the agent's decision(s), so that they can adjust their behavior accordingly.
\end{itemize}

\subsection{PQoS Inputs} 
\label{sub:pqos_inputs}

For the RAN-AI entity to operate properly, the availability of measurements (or ``inputs'') from diverse sources is the key.
The RAN-AI may have access to the following data from both the environment and the RAN~itself.
\begin{itemize}

\item  \emph{Context information.} This incorporates information about (i) the operational scenario, (ii) road elements, (iii) the network deployment, and (iv) the time of the day.

\item \emph{Trajectory of the users.} 
Driving applications may provide to the network the planned trajectory of end users, 
Moreover, the RAN-AI entity shall acquire data (based on trajectory statistics) about the gNB at which end users are/will be attached in future locations, as an indication of the cell load.

\item \emph{Traffic information.} 
The RAN-AI entity may gather traffic conditions from external control centers, which may also provide traffic predictions based on historical~data.

\item \emph{Network metrics.} 
The RAN-AI entity can get access to measurements gathered at the \gls{phy} and \gls{mac} layers (e.g., 
L1 measurements such as \gls{rsrp}, \gls{rsrq}, \gls{sinr}, \glspl{prb} utilization, and \gls{mcs} index),
\gls{rlc} and \gls{pdcp} layers (e.g., statistics of the data traffic exchanged across the users). 

\item \emph{Higher layers metrics.} 
The RAN-AI may be informed by the end users' applications about the experienced \gls{e2e} performance (e.g., mean, standard deviation, minimum and maximum value of delay and throughput).
\end{itemize}

\subsection{PQoS Countermeasures} 
\label{sub:pqos_countermeasures}

If QoS requirements are not satisfied, the role of the RAN-AI is to convert PQoS inputs into appropriate network countermeasures (or ``actions''). 
Besides operating directly on the driving patterns, 
the RAN-AI may undertake lower-layer actions (e.g., changing scheduling decisions, adapting the radio resource allocation as a function of the propagation conditions, modulating traffic requests based on the available network capacity, or modifying the system numerology to provide more resilient communication channels). 

Another action that can alleviate the burden on the channel is to reduce the size of the data generated at the application layer before transmission.
In our scenario, end users transmit \gls{lidar} data in the form of point clouds, whose size depends on the \emph{application mode}, i.e., the level of compression of the observations~\cite{nardo2022point}. 
In the context of this work, the RAN-AI entity implements an \gls{rl} agent whose goal is to identify the optimal application mode (i.e., the action) for the end users when generating the data (and thus the corresponding size of the packets to send). 
Finally, the RAN-AI communicates to each end user the corresponding best action that the agent has identified. This involves the transmission of a notification packet, which may be exposed to both transmission delays and errors. 

\section{Reinforcement Learning Model} 
\label{sub:rl_for_pqos}

\Acrfull{rl} is a powerful paradigm that models the target scenario as a \gls{mdp}, where time is discretized into slots and, in each slot~$t$, an agent observes the system state, takes a new action, and receives a reward or a penalty accordingly.
The goal of the agent is then to determine the optimal policy $\pi^*$, i.e., the map between states and actions $\pi: \mathcal{S} \rightarrow \mathcal{A}$ leading to maximizing the cumulative reward $G_t = \sum_{\tau=t}^{\infty} \lambda^{\tau-t} R(\tau)$, where $R$ is the reward and $\lambda \in [0,1)$ is the future reward discount factor. 

In this work, the RAN-AI is installed at the gNB and implements an \gls{rl} agent which controls all the vehicles connected to that cell.
Hence, the agent's actions correspond to the application modes described in Sec.~\ref{sub:pqos_countermeasures}, and the reward is obtained by evaluating the vehicles' performance.
This approach ensures that the size of the state/action spaces is invariant with respect to the number of users, and improves the learning efficiency since, during the training phase, the system can exploit the data gathered by all the vehicles.

\subsection{A Learning Framework for PQoS}
\label{sub:rl_framework}

In our system, at each step $t$, the agent observes the state $s_t \in \mathcal{S}$ of each vehicle, where $s_t$ is a vector containing the RAN-AI entity's inputs.
Given the state $s_t$, the agent computes the \emph{q-value} $Q(s_t,a)$ associated with each action $a \in \mathcal{A}$, where $Q(s_t,a)$ is the estimate of the cumulative reward $G_t$ that can be obtained playing action $a$ in state $s_t$, and then following the optimal policy.
During the training, the agent will eventually converge towards the action with the highest q-value, which ensures the best cumulative reward.

We adopt a \gls{drl} approach, and approximate the agent's policy by means of a \gls{nnet}, which makes it possible to handle continuous state spaces and overcome the ``curse of dimensionality'' phenomenon~\cite{bellman1966dynamic}.
In particular, we consider a Feed-Forward \gls{nnet}, with $S$ inputs and $A$ output neurons, and implement the \gls{relu} activation function across the different layers~\cite{agarap2018deep}. 
The input size ($S$) of the \gls{nnet} coincides with the number of input parameters of the RAN-AI entity, while the output size ($A$) corresponds to the number of possible agent's actions, i.e., the different application modes.
Hence, our architecture is trained to approximate the function $Q(\cdot): \mathcal{S} \times \mathcal{A} \rightarrow \mathbb{R}$, that establishes the quality of each state-action pair. 
In particular, the training of the agent is performed according to the {\gls{dql}} algorithm described in \cite{van2016deep}, which is an extended version of the classical {Q-learning} algorithm introduced in \cite{watkins1992q}.
The main details of our \gls{nnet} architecture are given in Table~\ref{tab:params}.

\subsection{Reward Function} 
\label{sub:reward_function}

The design of the reward function is a particularly critical task in \gls{rl}.
If the reward does not fully represent the system's requirements, the agent may learn an undesirable behavior, and lean towards suboptimal actions.
In our scenario, analyzing the performance of the system is not straightforward, as it depends on two different factors: 
\begin{itemize}
	\item The \acrfull{qos}: The agent's decision should ensure that the vehicle satisfies the agreed \gls{qos} in terms of different \glspl{kpi}, in particular in terms of maximum end-to-end delay $\delta_m$ and \gls{prr} (among the most representative KPIs for teleoperated driving scenarios~\cite{3GPP_22186}).
	
	\item The \acrfull{qoe}: The agent's decision should ensure that the quality of the transmitted data is good enough to perform teleoperated driving operations (e.g., object detection~\cite{rossi2021role}). This is measured based on the symmetric point-to-point Chamfer Distance $\mathrm{CD}_{\rm sym}$ between the transmitted data $\hat{D}$ and original data $D$ acquired by the LiDAR, expressed as~\cite{varischio2021hybrid}
	\medmuskip=0mu
	\thickmuskip=0mu
	\begin{equation}
	\begin{aligned}
	\mathrm{CD}_{\rm sym} = \sum_{\forall \mathbf{d}\in D} \min_{\forall \hat{\mathbf{d}}\in\hat{D}} \lVert \mathbf{\mathbf{d}} - \hat{\mathbf{d}} \rVert_2^2 + \sum_{\forall \hat{\mathbf{d}}\in \hat{D}} \min_{\forall \mathbf{d}\in D} \lVert \mathbf{ \mathbf{d}} - \hat{\mathbf{d}} \rVert_2^2.
	\label{eq:chamfer}
	\end{aligned}
	\end{equation}
	\medmuskip=6mu
	\thickmuskip=6mu
\end{itemize}

To make the agent capture both the above aspects, we design the reward as a piece-wise function, which returns $0$ whenever the \gls{qos} requirements are not met; otherwise, a positive value that depends on both the \gls{qos} and the \gls{qoe} of the end users.
Let $\hat{\text{PRR}}_t$, $\hat{\delta}_t$, and $\delta_m$ be the PRR and average delay of the vehicle at time $t$, and the maximum delay tolerated by the system (based on the use case of interest), respectively.
At time slot $t$, if the \gls{qos} requirements are met, i.e., $\hat{\delta}_t < \delta_m$ and $\hat{\text{PRR}}_t = 1$, the agent reward $R(t)$ is given by
\begin{equation}
\label{eq:reward}
R(t) =  (1-\alpha) \frac{\delta_m - \hat{\delta}_t}{\delta_m} + \alpha \frac{\mathrm{CD}_{\text{sym},m} - \hat{\mathrm{CD}}_{\text{sym},t}}{\mathrm{CD}_{\text{sym},m}},
\end{equation}
and is equal to $0$ otherwise. 
Specifically, in \eqref{eq:reward}, $\alpha$ is a positive value in $[0,1]$, while $\hat{\mathrm{CD}}_{\text{sym},t}$ and $\mathrm{CD}_{\text{sym},m}$ are the Chamfer Distance measured at time $t$ and the maximum Chamfer Distance that can be tolerated,~respectively.
The balancing between \gls{qos} and \gls{qoe} is determined by $\alpha$, which is a tuning parameter to be set according to the target~scenario.

\section{Performance Evaluation} 
\label{sec:performance_evaluation}

\begin{table}[t!]
\centering
\scriptsize
\renewcommand{\arraystretch}{1.3}
\caption{Simulation parameters.}
\label{tab:params}
\begin{tabular}{c|c|c}
    \toprule
    Parameter & Description & Value \\
    \hline
    
    $f_c$ & Carrier frequency & 3.5 GHz       \\ \hline
	$B$ & Total bandwidth                           & 50 MHz        \\ \hline
	$P_{TX}$ & Transmission power & 23 dBm        \\ \hline
	$T$ & RAN-AI update periodicity        & 100 ms        \\ \hline
	$\tau_s$ & Simulation time             & 80 s          \\ \hline
	$N_u$ & Number of vehicles & $\{1,\,5\}$ \\ \hline
	$\lambda$ & Discount factor  & 0.95          \\ \hline
	$\zeta$ & Learning rate      & $10^{-4}$     \\ \hline
	$\epsilon$ & Weight decay    & $10^{-3}$     \\ \hline
	$\alpha$ & QoS/QoE weight    		& $\{0.5,\, 1\}$ \\ \hline
	$\delta_m$ & Max. tolerated delay    & $50$ ms     \\ \hline
	$\mathrm{CD}_{\text{sym}, m}$ & Max. tolerated Chamfer Distance  & 45     \\ \hline
  \multicolumn{2}{c|}{Layer size (inputs $\times$ outputs)} & $8\times 12 \rightarrow 12\times 6 \rightarrow 6\times 3$ \\ 
  \bottomrule
	\end{tabular}
\end{table}

\begin{table*}[t!]
\centering
\footnotesize
\renewcommand{\arraystretch}{1.3}
\caption{Application modes and RL reward parameters.}
    \label{tab:reward}
\begin{tabular}{c|c|c|c|c|cc}
\toprule
\multirow{2}{*}{Application Mode} & \multirow{2}{*}{RangeNet++} & \multicolumn{2}{c|}{Draco}       & \multirow{2}{*}{Avg. file size [KB]} & \multirow{2}{*}{Chamfer Distance $\mathrm{CD}_{\rm sym}$} \\ \cline{3-4}
                        &                               & Compression & Quantization &                  &                  \\ \hline
 0 (baseline)                     & NO                            & \multicolumn{2}{c|}{NO}                & 3200      & 0        \\ \hline
 1450                  & NO                             & 5 & 14                &  200 & 0.000044  \\ \hline
 1451                  & LEVEL 1                             & 5 & 14                &  104 & 5.476881  \\ \hline
 1452                  & LEVEL 2                             & 5 & 14                &  17 & 35.634660  \\ \bottomrule
 
\end{tabular}
\end{table*}

In Sec.~\ref{sub:simulation_parameters_and_setup} we describe our simulation setup, 
while in Sec.~\ref{sub:numerical_results} we validate through numerical simulations the performance of our RAN-AI implementation for PQoS, compared to other baseline methods.

\subsection{Simulation Setup and Parameters} 
\label{sub:simulation_parameters_and_setup}
Our results are based on ns-3 simulations, thereby enabling full-stack end-to-end analyses.
To do so, we extended the \texttt{ns-3-mmwave} module~\cite{mezzavilla2018end} to incorporate a new RAN-AI class and its functionalities.
Simulation parameters are reported below and summarized in Table~\ref{tab:params}.

\paragraph{Scenario} 
Our PQoS framework was validated in a test scenario with one gNB covering a portion of the city of Bologna, Italy, and  $N_u=\{1,5\}$ vehicles moving according to realistic mobility traces generated using \gls{sumo}~\cite{SUMO2012}. 
The system is operating at 3.5 GHz (corresponding to NR band n78) with a bandwidth of 50 MHz. 

\paragraph{V2X application} 
Each vehicle runs an uplink application streaming LiDAR data, modeled according to the Kitti multi-modal dataset~\cite{kitti-dataset}. Moreover, it receives in the downlink commands from the remote driver for teleoperated driving operations. Each raw LiDAR perception generates a point cloud of around 120\,000 points at 10 Hz, with an average file size of  3\,200 KB.
Based on~\cite{varischio2021hybrid}, compression is accomplished using Draco~\cite{Draco} (a software designed by Google to compress 3D-like data) combined with the semantic segmentation functionalities of RangeNet++~\cite{rnet} (a \gls{nnet} able to assign class labels to data points). 
Our compression pipeline consists of the following~steps:
\begin{itemize}
  \item We first infer semantic segmentation of LiDAR data with RangeNet++, so as to identify the most valuable objects in the scene. Three semantic levels are defined:
  \begin{itemize}
    \item LEVEL 0: The raw LiDAR acquisition is considered. 
    \item LEVEL 1: The points associated to the road elements are removed from the cloud, thus reducing the file size. 
    \item LEVEL 2: The points associated to buildings, vegetation, and traffic signs are also removed; the resulting data consists only of dynamic elements like pedestrians and vehicles, i.e., the most critical elements in autonomous and remote driving scenarios.
  \end{itemize}
  \item The resulting point cloud is then compressed with Draco, which defines 15 quantization levels and 11 compression levels that trade off efficiency against~speed.
\end{itemize}
We consider four application modes, as reported in Table~\ref{tab:reward}. 

\paragraph{RL agent architecture} 

The state of our RL agent includes only a subset of the RAN-AI input parameters (Sec.~\ref{sub:pqos_inputs}), to reduce the state space dimension, and ensure a faster learning convergence. Specifically, 
we focused on RAN-level metrics usually available at the gNB, averaged over a reporting period of 100~ms, i.e., the value of the \gls{mcs}, the number of \gls{ofdm} symbols used to transmit, the value of the \gls{sinr}, the mean/max/min/std of the packets' delay, and the \gls{prr} at the \gls{pdcp} layer.
We also limited the agent's action space to three actions, corresponding to application modes $\{1450$, $1451$, $1452\}$,\footnote{From preliminary experiments, we obtained that application mode 0 (where data are not compress nor segmented) was never selected by the agent during training, so it was not included in the action space to promote faster convergence; this strategy will represent the benchmark solution in our tests.} so that $S=8$ and $A=3$.

With respect to \gls{dql}, we set the discount factor of the \gls{rl} algorithm to $\lambda=0.95$ and we adopt a batch-learning approach, where the learning transitions are grouped in batches of size $B_{\text{size}}=10$. 
We implement the \gls{adam} algorithm to update the weights of the \gls{nnet}, considering $\zeta=10^{-5}$ as the maximum learning rate, and $\epsilon=10^{-3}$ as the weight decay~\cite{kingma2014adam}.
The agent's training follows two subsequent phases, organized into episodes of $800$ steps each, where the step duration is set to $100$ ms so that each episode lasts $80$ seconds.
First, we perform an \emph{offline} training phase, where the agent's actions are kept fixed for the whole duration of the episode.
Then, we perform an \emph{online} training phase, where the agent's actions change in time according to an \emph{$\epsilon$-greedy} exploration policy. 
The duration of each training phase depends on the number of vehicles in the scenario, and is set to $2\,500$ when $N_u=1$ and $500$ when $N_u=5$.

In terms of the reward function in Eq.~\eqref{eq:reward}, we set $\delta_m=50$ ms, as specified by the 5GAA for teleoperated driving applications~\cite{5gaa2020cv2x}, and $\mathrm{CD}_{\text{sym}, m}=45$, while the Chamfer Distance $\mathrm{CD}_{\text{sym}}$ associated to each application mode is reported in Table~\ref{tab:reward}. Finally, we set $\alpha\in\{0.5,1\}$ to consider different QoS/QoE weights.

\paragraph{Performance evaluation} 
\label{par:performance_evaluation}
To validate our framework, we compared the following action policies:
\begin{itemize}
  \item \emph{DQL} (proposed), where at each step the agent implements our proposed RL framework described in Sec.~\ref{sub:rl_for_pqos};
  \item \emph{Constant} (benchmark), where at the beginning of the simulation the end user mantains one application mode among $\{0,$ $1450$, $1451$, $1452\}$ for the whole simulation. 
\end{itemize}
The two strategies have been tested separately, and will be compared in terms of (i) the reward gained by the agent, (ii) the probability to satisfy the \gls{qos} requirements, (iii) the user's \gls{qos}, expressed in terms of delay at the application layer, and (iv) the user's \gls{qoe}, expressed in terms of $\mathrm{CD}_{\rm sym}$.

\subsection{Numerical Results} 
\label{sub:numerical_results}

\begin{figure*}[h!]
	\centering
	\begin{subfigure}{.65\columnwidth}
		\centering
		\includegraphics[width=\textwidth]{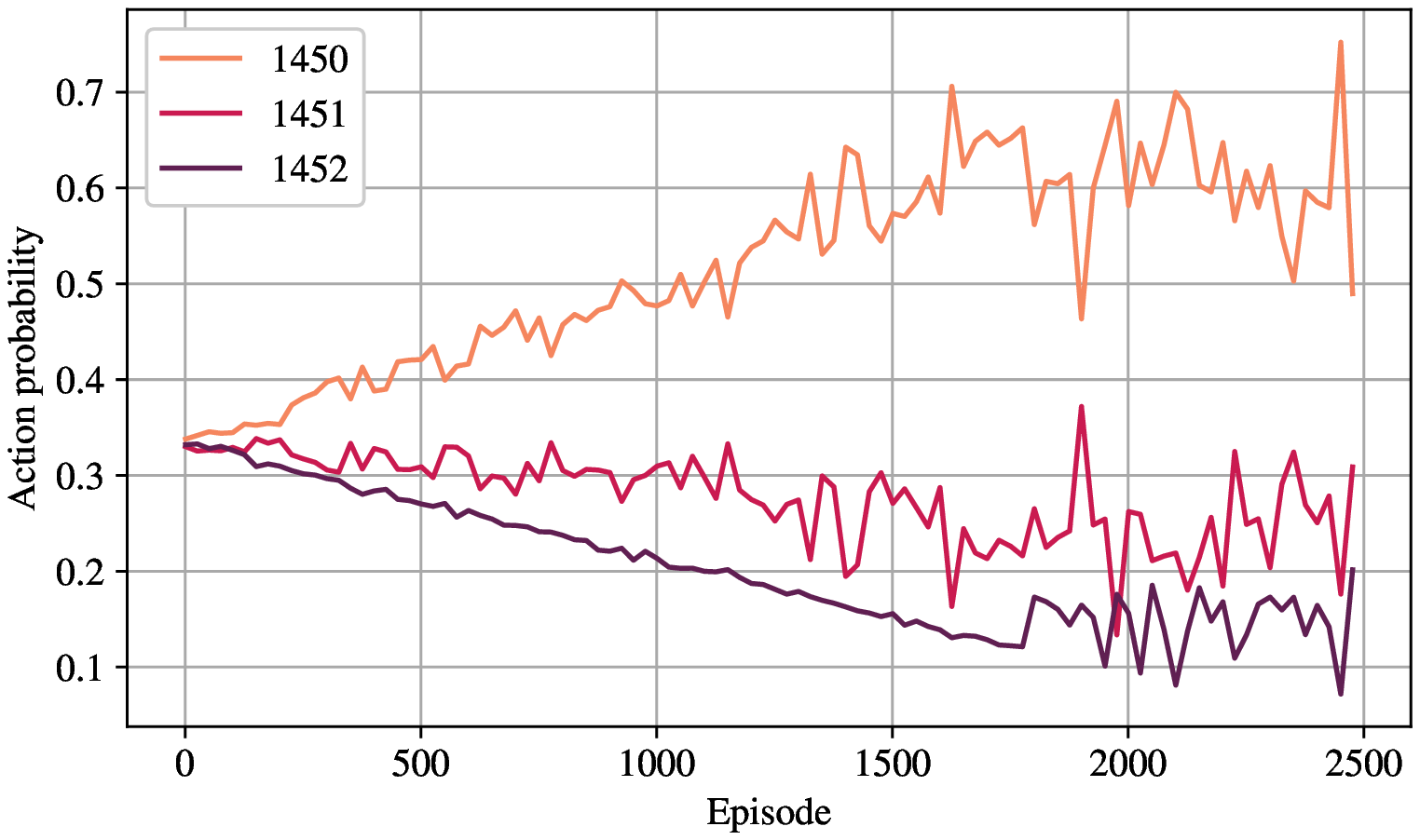}
		\caption{$N_u=1$, $\alpha=1$.}
		\label{fig:user_1_train}
	\end{subfigure}\quad
	\begin{subfigure}{.65\columnwidth}
		\centering
		\includegraphics[width=\textwidth]{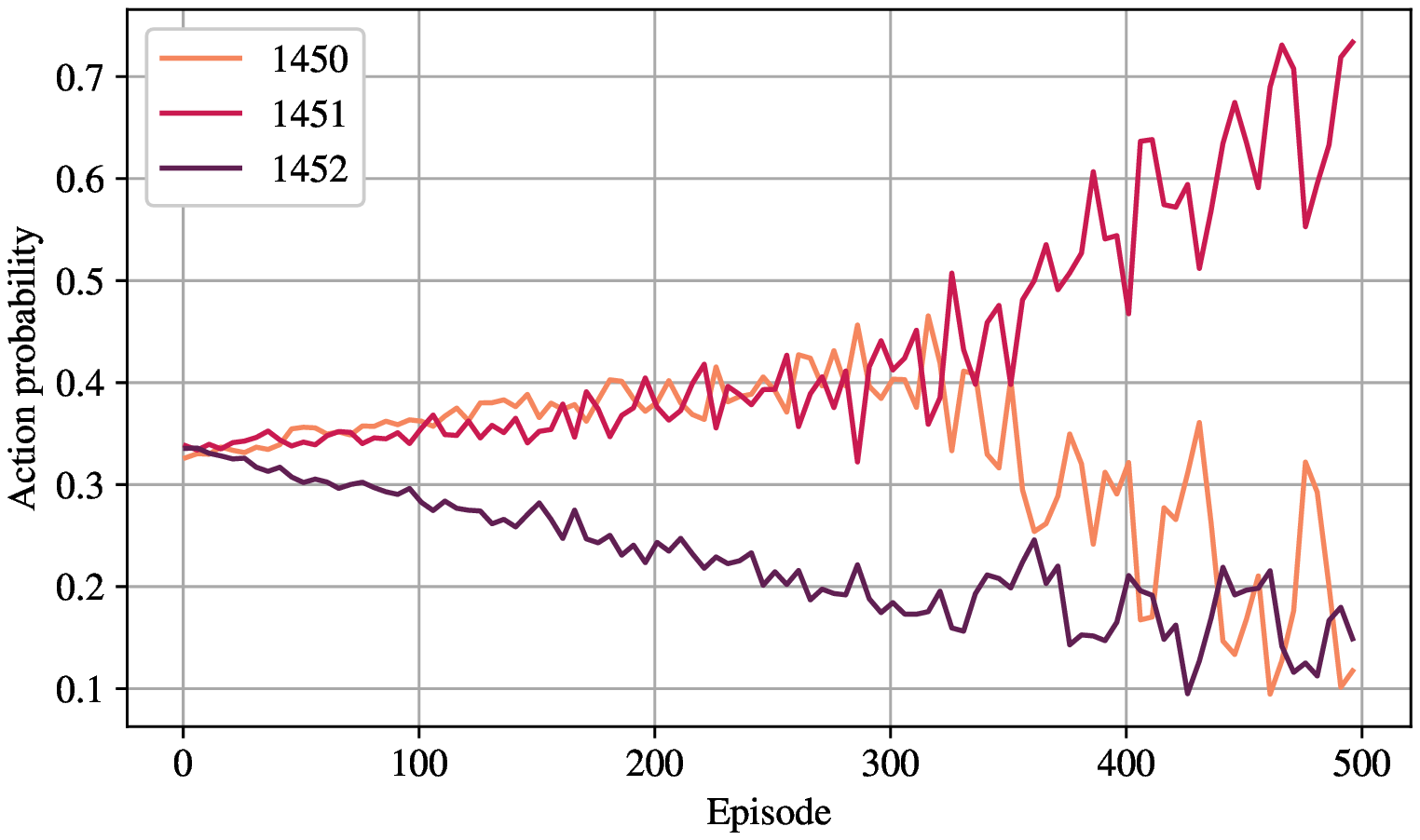}
		\caption{$N_u=5$, $\alpha=1$.}
		\label{fig:user_5_train}
	\end{subfigure} \quad
		\begin{subfigure}{.65\columnwidth}
		\centering
		\includegraphics[width=\textwidth]{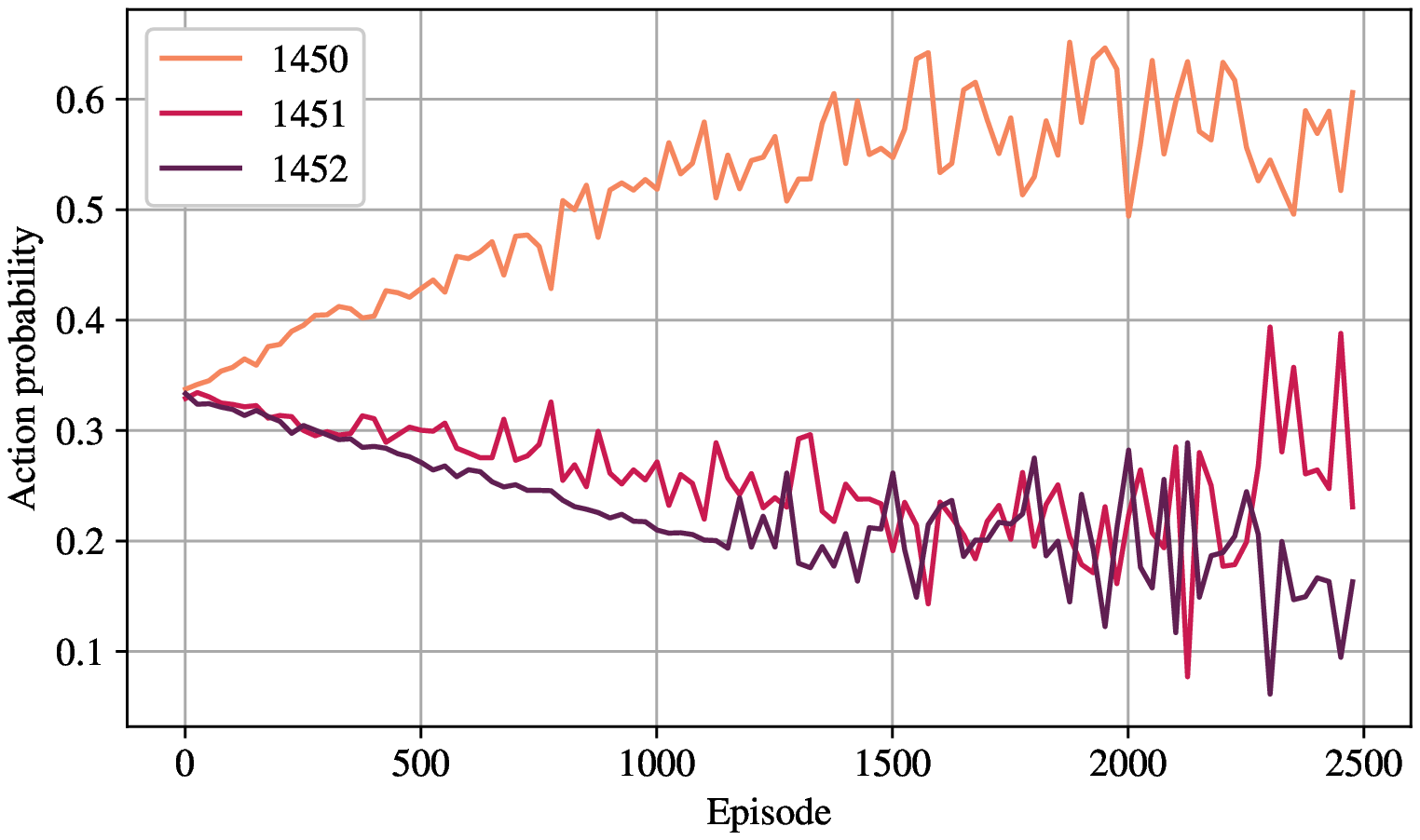}
		\caption{$N_u=1$, $\alpha=0.5$.}
		\label{fig:user_1_train_05}
	\end{subfigure}
	\caption{Action probability during the training phase, vs. the number of vehicles $N_u$ and $\alpha$.}
	\label{fig:agent_train}
    \vspace{-3.75ex} 
\end{figure*}

First, in Fig.~\ref{fig:agent_train} we evaluate the DQL statistics of the training phase vs. $N_u$ and $\alpha$. 
We observe that, during the first phase of the training period, the agent mostly makes random decisions because of the $\epsilon$-greedy policy, which ensures that the action and state spaces are fully explored.
As the training progresses, the agent starts acting greedily and prioritizes the actions that maximize the long-term reward. 
In this case, the agent's decisions depend on the observed state and, consequently, the action probability distribution may change completely from an episode to another, which justifies the variability in Fig.~\ref{fig:agent_train}.
We can see that, when $N_u=1$, the agent prefers application mode $1450$, since it offers the best trade-off between QoS and QoE compared to other actions.
Instead, when $N_u=5$, the system is more congested, and the agent tends to penalize more the actions leading to a violation of the QoS requirements; as such, action $1451$ (where the original LiDAR data is subject to both compression and segmentation before transmission) is preferred most of the time. 

\begin{figure}[t!]
	\centering
	\setlength{\belowcaptionskip}{-0.63cm}
	\includegraphics[width=0.8\columnwidth]{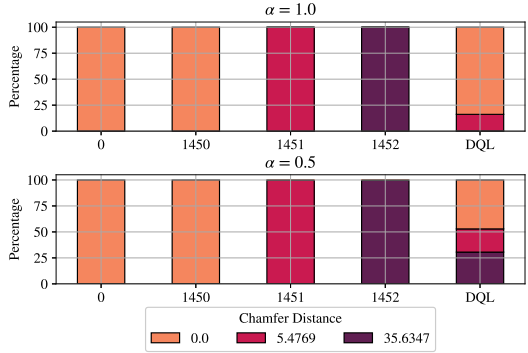}
	\caption{Chamfer Distance ($\mathrm{CD}_{\text{sym}}$) distribution, for $N_u=1$.}
	\label{fig:cd_user_1}
\end{figure}

\begin{figure}[t!]
	\centering
	\setlength{\belowcaptionskip}{-0.43cm}
	\includegraphics[width=0.8\columnwidth]{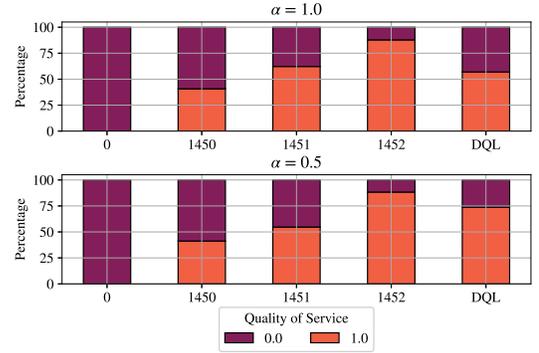}
	\caption{QoS distribution, when for $N_u=1$.}
	\label{fig:qos_user_1}
\end{figure}

Second, we analyze the QoE and QoS performance of the \gls{dql} agent against other benchmarks in different system configurations, during a test phase of $100$ episodes.  
In Fig.~\ref{fig:cd_user_1}, we illustrate the distribution of the Chamfer Distance (an indication of the QoE of the system), considering $N_u=1$ and for $\alpha\in\{0.5, 1.0\}$. 
We observe that, when the RAN-AI entity implements strategies $\{0,$ $1450$, $1451$, $1452\}$, the Chamfer Distance remains constant throughout all the episodes.
Instead, with DQL, the RAN-AI entity tries to adapt to the conditions of the environment.
In particular, when $\alpha=1$, the agent has an incentive to improve the \gls{qoe}'s reward component, and prioritizes action $1450$, so that ${\mathrm{CD}}_{\text{sym}}$ is $0$ for more than $45\%$ of the time.
On the other hand, for $\alpha=0.5$, the reward function discourages an overly aggressive behavior, which may lead to a violation of the \gls{qos} requirements, and prefers action $1452$, which however leads to some QoE~degradation.
Similarly, Fig.~\ref{fig:qos_user_1} plots the distribution of the \gls{qos} for the different strategies, where a $\text{QoS}$ of $0$ implies that the delay/PRR requirements of the teleoprated driving application are not satisfied.
We observe that \gls{dql} improves as $\alpha$ decreases, since the agent receives a higher reward when the delay is minimized.
In particular, when $\alpha=1$ ($\alpha=0.5)$, the QoS requirements are satisfied around $50\%$ ($75\%$) of the time.
Notably, while the Constant $1452$ action ensures better QoS (since the data are compressed and segmented before transmission, which can reduce the size of the packets to send), it is characterized by a very high ${\mathrm{CD}}_{\text{sym}}$ (Fig.~\ref{fig:cd_user_1}) and, consequently, a bad reward distribution. 
Similarly, while the Constant $1450$ action promotes better QoE, it results in a QoS degradation of up to $40\%$ compared to DQL when $\alpha=0.5$.

\begin{figure}[t!]
	\centering
	\setlength{\belowcaptionskip}{-0.53cm}
	\includegraphics[width=0.8\columnwidth]{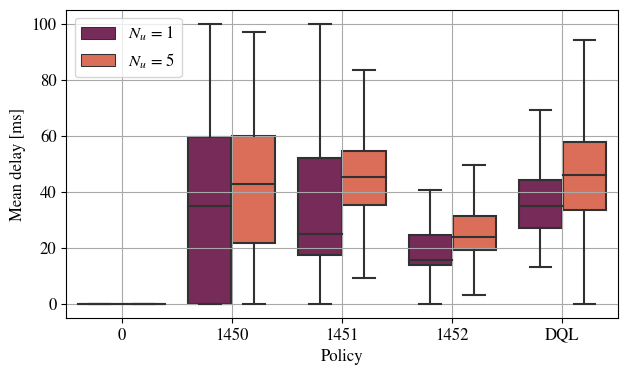}
	\caption{Mean delay (application layer), for $\alpha=1$.}
	\label{fig:delay_user_1}
\end{figure}

The above considerations are validated in Fig.~\ref{fig:delay_user_1}, which reports the distribution of the mean packet delay at the application layer, i.e., the distribution of the average delay of all the packets delivered at the application layer during the same time slot.  
Specifically, we use the box plot representation, where the black line in the middle of each box is the median, the edges represent the 25th and 75th percentiles, while the whiskers identify the outliers of the distribution.
We observe that all the policies can maintain the median delay below the required threshold ($\delta_m=50$ ms) for both $N_u=1$ and $N_u=5$.
However, while for Constant $1450$ and $1451$ the top whiskers of the distribution approach a value of $100$ ms, which indicates that such strategies may be unable to meet the \gls{qos} requirements when the working conditions are critical, for DQL the whiskers are equal to $70$ and $95$ ms for $N_u=1$ and $N_u=5$, respectively.
Notice that the only strategy with a better QoS is Constant $1452$, resulting~in a delay distribution squeezed towards zero, thus ensuring that QoS requirements are rarely violated, at the expense of a poor~QoE. 

\begin{figure}[t!]
	\centering
	\begin{subfigure}{\columnwidth}
		\centering
		\includegraphics[width=0.7\columnwidth]{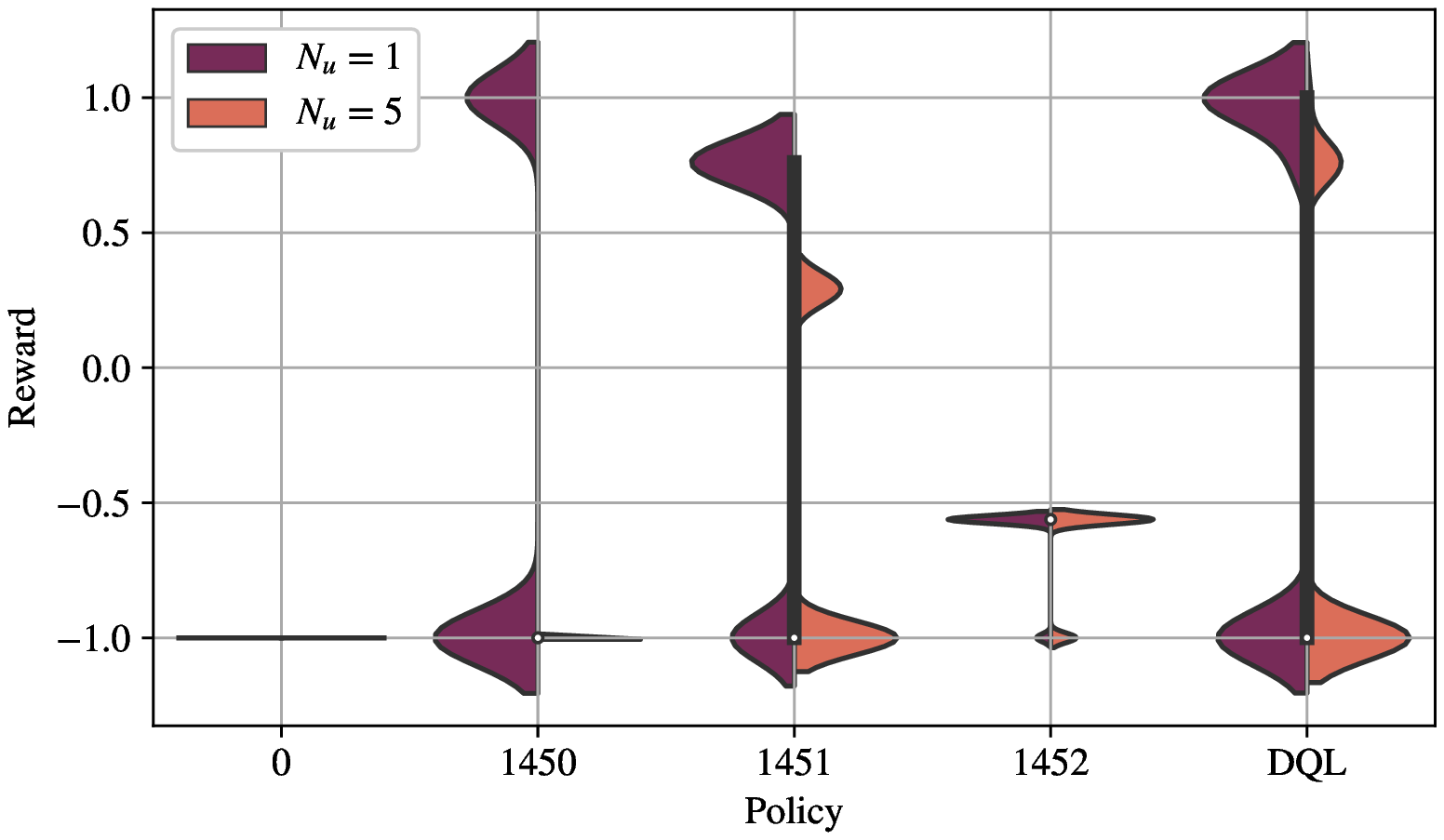}
		\caption{$\alpha=1.0$  vs. $N_u$.\vspace{0.33cm}}
		\label{fig:reward_alpha_1}
	\end{subfigure}
	\begin{subfigure}{\columnwidth}
		\centering
		\includegraphics[width=0.7\columnwidth]{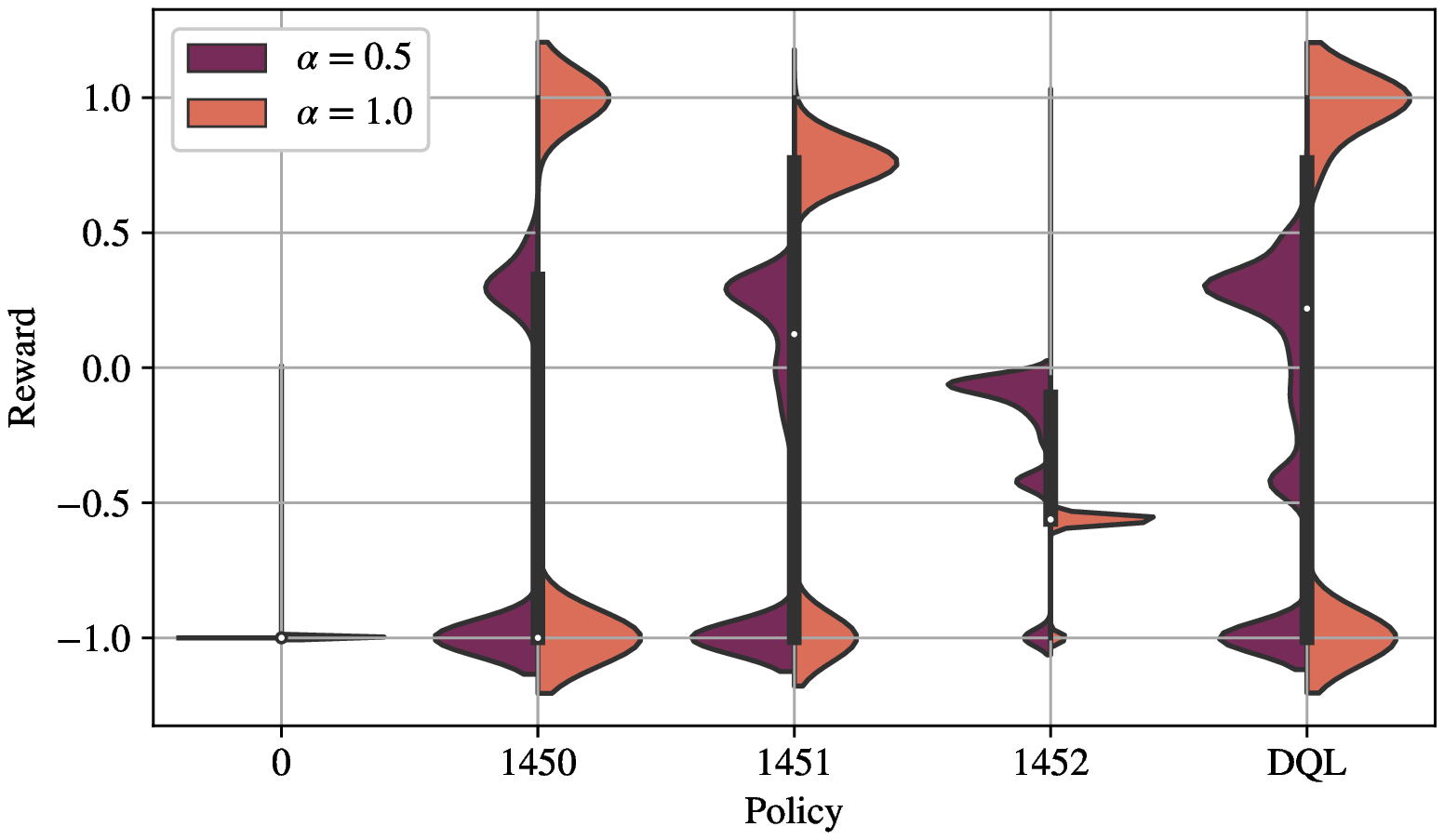}
		\caption{$N_u=1$ vs. $\alpha$.}
		\label{fig:reward_vehicle_1}
	\end{subfigure}
	\caption{Distribution of the agent's reward, normalized in $[-1, +1]$.\vspace{-0.33cm}}
	\label{fig:agent_reward}
\end{figure}

To evaluate the trade-off between QoE and QoS of the different PQoS policies, in Fig.~\ref{fig:agent_reward} we plot the agent's reward (normalized in $[-1, +1]$) for different values of $\alpha$ and $N_u$. Specifically (i) the white dot represents the median, (ii) the thick black bar in the center represents the interquartile range, (iii) the thin black line represents the rest of the distribution, except for ``outliers.'' Wider sections of the violin plot indicate values that occur more frequently.
From Fig.~\ref{fig:reward_alpha_1} we can see that the \gls{dql} solution achieves the best trade-off between QoS and QoE, meaning that the adaptive behavior of DQL is desirable for PQoS. We notice that the Constant $0$ and $1452$ policies display the lowest reward due to their poor QoS and QoE performance, respectively. 
For $N_u=1$, DQL and the Constant $1450$ and $1451$ policies achieve similar rewards.
On the other hand, when $N_u=5$, i.e., considering more congested networks, the advantages of \gls{dql} are more evident. 
Notably, \gls{dql} is the only strategy that can provide a reward higher than $0.5$, while for Constant $1450$ and $1451$ the reward peaks at about $-1$ and $0.4$, respectively. 
By tuning parameter $\alpha$, it is possible to further adjust the policy learned by the DQL agent, promoting more or less conservative communication settings. From Fig.~\ref{fig:reward_vehicle_1}, we observe that with $\alpha=0.5$ the DQL agent is able to reduce the probability of negative rewards, compared to $\alpha=1$. In this case, however, the maximum reward is $0.3$ (vs. $1$ when $\alpha=1$), through still higher than any competitor.

\section{Conclusions and Future Works} 
\label{sec:conclusions_and_future_works}
In this paper we analyzed the potential of PQoS to predict and optimize autonomous networks. Specifically, we proposed the design of a new ``RAN-AI'' entity, installed in the gNB and interacting with a custom RL agent, to identify the optimal set of PQoS actions/countermeasures to satisfy QoS requirements of end users. We performed simulations in ns-3 in a teleoperated driving scenario, and demonstrated that the adaptive behavior of our RL model can achieve the best trade-off between QoS and QoE performance, compared to other baseline solutions that do no support machine learning. 
Also, we make the case that, by properly tuning the parameters of the reward function, it is possible to adjust the policy learned by the learning agent, promoting more or less conservative communication settings.

Among our future research activities, we will extend our current PQoS framework to incorporate advanced functionalities, including the support for multi-cell scenarios, as well as the definition of new ML technologies and countermeasures for PQoS (e.g., based on federated learning). In particular, we will investigate whether a fully distributed architecture, in which end machines make autonomous decisions, can promote more efficient PQoS.

\bibliographystyle{IEEEtran}
\bibliography{bibl.bib}

\begin{thebibliography}{10}
\providecommand{\url}[1]{#1}
\csname url@samestyle\endcsname
\providecommand{\newblock}{\relax}
\providecommand{\bibinfo}[2]{#2}
\providecommand{\BIBentrySTDinterwordspacing}{\spaceskip=0pt\relax}
\providecommand{\BIBentryALTinterwordstretchfactor}{4}
\providecommand{\BIBentryALTinterwordspacing}{\spaceskip=\fontdimen2\font plus
\BIBentryALTinterwordstretchfactor\fontdimen3\font minus
  \fontdimen4\font\relax}
\providecommand{\BIBforeignlanguage}[2]{{%
\expandafter\ifx\csname l@#1\endcsname\relax
\typeout{** WARNING: IEEEtran.bst: No hyphenation pattern has been}%
\typeout{** loaded for the language `#1'. Using the pattern for}%
\typeout{** the default language instead.}%
\else
\language=\csname l@#1\endcsname
\fi
#2}}
\providecommand{\BIBdecl}{\relax}
\BIBdecl

\bibitem{shariatmadari2015machine}
H.~{Shariatmadari}, R.~{Ratasuk}, S.~{Iraji}, A.~{Laya}, T.~{Taleb},
  R.~{Jäntti}, and A.~{Ghosh}, ``{Machine-type communications: current status
  and future perspectives toward 5G systems},'' \emph{IEEE Communications
  Magazine}, vol.~53, no.~9, pp. 10--17, Sep. 2015.

\bibitem{boban2021predictive}
M.~Boban, M.~Giordani, and M.~Zorzi, ``{Predictive Quality of Service (PQoS):
  The Next Frontier for Fully Autonomous Systems},'' \emph{IEEE Network}, 2021.

\bibitem{giordani2020toward}
M.~Giordani, M.~Polese, M.~Mezzavilla, S.~Rangan, and M.~Zorzi, ``{Toward 6G
  networks: Use cases and technologies},'' \emph{IEEE Communications Magazine},
  vol.~58, no.~3, pp. 55--61, Mar. 2020.

\bibitem{Moral:1996}
P.~Del~Moral, ``Nonlinear filtering: Interacting particle resolution,''
  \emph{Comptes Rendus de l'Acad{\'e}mie des Sciences-Series I-Mathematics},
  vol. 325, no.~6, pp. 653--658, Sep. 1997.

\bibitem{Wan:2000}
E.~A. Wan and R.~V.~D. Merwe, ``{The unscented Kalman filter for nonlinear
  estimation},'' in \emph{IEEE Adaptive Systems for Signal Processing,
  Communications, and Control Symposium}, Oct. 2000, pp. 153--158.

\bibitem{mason2020adaptive}
F.~Mason, M.~Giordani, F.~Chiariotti, A.~Zanella, and M.~Zorzi, ``An adaptive
  broadcasting strategy for efficient dynamic mapping in vehicular networks,''
  \emph{IEEE Transactions on Wireless Communications}, vol.~19, no.~8, pp.
  5605--5620, Aug. 2020.

\bibitem{5gaa2019qos}
{5GAA Automotive Association}, ``{Making 5G Proactive and Predictive for the
  Automotive Industry},'' \emph{White Paper}, Aug. 2019.

\bibitem{carron2016machine}
A.~Carron, M.~Todescato, R.~Carli, L.~Schenato, and G.~Pillonetto, ``{Machine
  learning meets Kalman filtering},'' in \emph{IEEE 55th Conference on Decision
  and Control (CDC)}, 2016.

\bibitem{wang2017machine}
M.~Wang, Y.~Cui, X.~Wang, S.~Xiao, and J.~Jiang, ``{Machine learning for
  networking: Workflow, advances and opportunities},'' \emph{IEEE Network},
  vol.~32, no.~2, pp. 92--99, Mar. 2017.

\bibitem{nardo2022point}
F.~Nardo, D.~Peressoni, P.~Testolina, M.~Giordani, and A.~Zanella, ``{Point
  Cloud Compression for Autonomous Driving: A Performance Comparison},'' in
  \emph{IEEE Wireless Communications and Networking Conference (WCNC) [To
  Appear]}, 2022.

\bibitem{bellman1966dynamic}
R.~Bellman, ``Dynamic programming,'' \emph{Science}, vol. 153, no. 3731, pp.
  34--37, Jul. 1966.

\bibitem{agarap2018deep}
A.~F. Agarap, ``{Deep learning using rectified linear units (ReLu)},''
  \emph{arXiv preprint arXiv:1803.08375}, 2018.

\bibitem{van2016deep}
H.~Van~Hasselt, A.~Guez, and D.~Silver, ``{Deep reinforcement learning with
  double Q-learning},'' in \emph{Proceedings of the AAAI conference on
  artificial intelligence}, vol.~30, no.~1, 2016.

\bibitem{watkins1992q}
C.~J. Watkins and P.~Dayan, ``Q-learning,'' \emph{Machine learning}, vol.~8,
  no. 3-4, pp. 279--292, May 1992.

\bibitem{3GPP_22186}
{3GPP}, ``{Service requirements for enhanced V2X scenarios (Release 15)},''
  \emph{TS 22.186}, Sep. 2018.

\bibitem{rossi2021role}
V.~Rossi, P.~Testolina, M.~Giordani, and M.~Zorzi, ``{On the Role of Sensor
  Fusion for Object Detection in Future Vehicular Networks},'' in \emph{Joint
  European Conference on Networks and Communications 6G Summit (EuCNC/6G
  Summit)}, 2021.

\bibitem{varischio2021hybrid}
A.~Varischio, F.~Mandruzzato, M.~Bullo, M.~Giordani, P.~Testolina, and
  M.~Zorzi, ``{Hybrid Point Cloud Semantic Compression for Automotive Sensors:
  A Performance Evaluation},'' \emph{IEEE International Conference on
  Communications (ICC)}, 2021.

\bibitem{mezzavilla2018end}
M.~{Mezzavilla}, M.~{Zhang}, M.~{Polese}, R.~{Ford}, S.~{Dutta}, S.~{Rangan},
  and M.~{Zorzi}, ``{End-to-End Simulation of 5G mmWave Networks},'' \emph{IEEE
  Communications Surveys and Tutorials}, vol.~20, no.~3, pp. 2237--2263, Apr.
  2018.

\bibitem{SUMO2012}
D.~Krajzewicz, J.~Erdmann, M.~Behrisch, and L.~Bieker, ``Recent development and
  applications of {SUMO - Simulation of Urban MObility},'' \emph{International
  Journal On Advances in Systems and Measurements}, vol.~5, no. 3\&4, pp.
  128--138, Dec. 2012.

\bibitem{kitti-dataset}
\BIBentryALTinterwordspacing
{Karlsruhe Institute of Technology and Toyota Technological Institute at
  Chicago}, ``{The KITTI Vision Benchmark Suite}.'' [Online]. Available:
  \url{http://www.cvlibs.net/datasets/kitti/}
\BIBentrySTDinterwordspacing

\bibitem{Draco}
Google, ``{Draco 3D Data Compression},'' 2017, [Online]. Available:
  https://github.com/google/draco.

\bibitem{rnet}
A.~Milioto, I.~Vizzo, J.~Behley, and C.~Stachniss, ``{RangeNet++: Fast and
  accurate LiDAR semantic segmentation},'' in \emph{IEEE/RSJ International
  Conference on Intelligent Robots and Systems (IROS)}, 2019.

\bibitem{kingma2014adam}
D.~P. Kingma and J.~Ba, ``Adam: A method for stochastic optimization,''
  \emph{arXiv preprint arXiv:1412.6980}, 2014.

\bibitem{5gaa2020cv2x}
{5GAA}, ``{C-V2X Use Cases Volume II: Examples and Service Level
  Requirements},'' \emph{White Paper}, Oct. 2020.

\end{thebibliography}

\end{document}